\documentstyle[epsfig]{article}
\def\cpc#1#2#3{Comp.\ Phys.\ Commun.\ #1 (19#3) #2}
\def\ib#1#2#3{ibid.\ #1 (19#3) #2}
\def\jpg#1#2#3{J.\ Phys.\ G#1 (19#3) #2}
\def\np#1#2#3{Nucl.\ Phys.\ B#1 (19#3) #2}
\def\pl#1#2#3{Phys.\ Lett.\ #1B (19#3) #2}
\def\pr#1#2#3{Phys.\ Rev.\ D#1 (19#3) #2}
\def\prl#1#2#3{Phys.\ Rev.\ Lett.\ #1 (19#3) #2}
\def\zp#1#2#3{Z.\ Phys.\ C#1 (19#3) #2}
\def\eref#1{Eq.~(\ref{#1})}
\def\erefs#1#2{Eqs.~(\ref{#1}) and (\ref{#2})}
\def\fref#1{Fig.~\ref{#1}}
\def\frefs#1#2{Figs.~\ref{#1} and \ref{#2}}
\def\sref#1{Sect.~\ref{#1}}
\def\srefs#1#2{Sects.~\ref{#1} and \ref{#2}}
\def\stub#1#2{#1_{\mbox{\tiny #2}}}
\def\aeff{\stub{\alpha}{eff}}
\def\beq{\begin{equation}}
\def\beeq{\begin{eqnarray}}
\def\eeq{\end{equation}}
\def\eeeq{\end{eqnarray}}
\def\as{\stub{\alpha}{S}}
\def\bm#1{{\mbox{\boldmath $#1$}}}
\def\kt2{{\mbox{\boldmath $k$}_\perp^2}}
\def\cF{{\cal F}}
\def\ee{e^+ e^-}
\def\eps{\epsilon}

\def\asp{\frac\as\pi}
\def\aptn{\alpha^{(n)}_{\mbox{\tiny pert}}}
\def\daen{\delta\alpha^{(n)}_{\mbox{\tiny eff}}}
\def\dae2{\delta\alpha^{(n=2)}_{\mbox{\tiny eff}}}
\def\Fper{\stub{F}{pert}}
\def\Fptn{F^{(n)}_{\mbox{\tiny pert}}}
\def\Fpwn{F^{(n)}_{\mbox{\tiny pow}}}
\def\mI{\stub{\mu}{I}}
\def\VEV#1{\langle{#1}\rangle}
\begin{document}

\title{Renormalon Phenomena in Jets and Hard Processes
\footnote{Talk at XXVII International Symposium on
Multiparticle Dynamics, Frascati, Italy, 8--12 September 1997.
Research supported in part by the U.K.\ Particle Physics
and Astronomy Research Council and by the EC Programme
``Training and Mobility of Researchers", Network
``Hadronic Physics with High Energy Electromagnetic Probes",
contract ERB FMRX-CT96-0008.}}

\author{B.R.\ Webber\\
Cavendish Laboratory, University of Cambridge,\\
Cambridge CB3 0HE, UK.}
\maketitle

\begin{abstract}
The `renormalon' or `dispersive' method for
estimating non-perturbative corrections to
QCD observables is reviewed. The corrections
are power-suppressed, i.e. of the form $A/Q^p$
where $Q$ is the hard process momentum scale.
The renormalon method exploits the connection between
divergences of the QCD perturbation series and
low-momentum dynamics to predict the power, $p$.
The further assumption of an approximately universal
low-energy effective strong coupling leads to relationships
between the coefficients $A$ for different observables.
Results on $1/Q^2$ corrections to deep inelastic
structure functions and $1/Q$ corrections
to event shapes are presented and compared with
experiment. Shape variables that could be free of
$1/Q$ and $\as(Q^2)/Q$ corrections are suggested. 
\end{abstract}

\renewcommand{\thefootnote}{\fnsymbol{footnote}}
\section{Introduction}\label{sec_intro}
The question of how best to compare QCD predictions with
strong interaction data is an old one.  The problem is that
we are usually comparing predictions
about quarks and gluons -- the so-called parton level -- with data
on mesons and baryons -- the hadron level.  We must then choose
carefully those quantities for which the differences between partons
and hadrons are not expected to be important.

Our criteria for choosing what to compare are in fact all based on
parton-level perturbation theory, so the argument looks in danger of
being circular.  However, we believe that the conversion of partons
into hadrons -- hadronization -- is a long-distance process. So we
can select those observables that are insensitive to long-distance
physics (infrared-safe quantities) or in which long-distance
contributions can be separated out (factorizable quantities).
These properties can be established entirely within perturbation
theory, and it seems reasonable to expect that they are also
true in the real world.

Increasingly, though, we need to know more quantitatively which
are the {\em best} QCD observables, with the least sensitivity
to long-distance physics.  One way to study this is to make a
model for the non-perturbative hadronization process and to
look at the size of the resulting hadronization corrections.
This has been a very fruitful approach, and is still the
standard procedure of most experiments, but it does have
its limitations.\footnote{As one of the engineers of
this approach, I feel free to criticise it.}
It doesn't apply to long-distance physics other than
hadronization, for example the parton distributions
inside colliding hadrons.  It's a `black-box' procedure
that often doesn't provide much insight, and has many
adjustable parameters that may conceal wrong physics
assumptions.  Perhaps most seriously of all, it doesn't
really apply in the circumstances in which it is normally
used. By that I mean that the `parton level' to which
the hadronization model is applied is a somewhat
mutilated version of the final states that appear
in a perturbative calculation. It contains approximate
matrix elements, infrared and collinear cutoffs, angular
ordering instead of genuine quantum interference, etc.
Therefore, even if the hadron-level predictions of the
model agree very well with experiment, as they often
do, we cannot be sure that the hadronization effects
are correctly estimated.

In the past two or three years an alternative method
has developed, based more directly on perturbation theory.
The approach is a logical extension of the identification
of suitable (infrared-safe or factorizable) observables
within perturbation theory. One now asks which amongst
these observables are expected to be more
sensitive to long-distance physics, and which less
sensitive. This can be established by seeing
whether low-momentum regions contribute much to
the perturbative prediction, even when the overall
momentum scale $Q$ is high.  Generally one finds that
the low-momentum contribution is of the form $A/Q^p$,
i.e., power-suppressed, modulo logarithms.  Simply
by examining Feynman graphs, one can compute the
powers $p$ for different observables, which is already
useful. More speculatively, one can try to
estimate the coefficients $A$, and to relate
them to the power corrections observed experimentally.
In the following Sections, I shall review recent work
on both of these aspects of the problem.

The approach outlined above has become known
as the `renormalon' or `dispersive' method for
estimating power corrections. \sref{sec_renorm}
explains the origin of this terminology and its
connection with low-momentum contributions. Then
\sref{sec_pow} shows how the power corrections
arise. Applications of the method to hadronic
structure functions and event shape variables
are reviewed in \srefs{sec_dis}{sec_shape} respectively.
Event shape variables are particularly `bad'
observables, in that they typically have the largest
power corrections, with $p=1$. In \sref{sec_better},
I discuss the possibilities for constructing better
shape variables, with $p>1$.  Finally \sref{sec_conc}
contains some conclusions and further speculations.

\section{Renormalons}\label{sec_renorm}
The usual starting point for a discussion of
renormalons \cite{renormalons} is the
observation that the QCD perturbation series
is not expected to converge.  There may indeed already
be signs of this in some of the quantities that have
been computed to order $\as^3$, for example the
Gross-Llewellyn Smith sum rule for deep inelastic neutrino
scattering:
\beeq\label{GLS}
\int_0^1 dx\, (F_3^\nu + F_3^{\bar\nu}) &=& 6\biggl[ 1-\asp
-3.6\left(\asp\right)^2\nonumber\\
& &-19\left(\asp\right)^3+\cdots\biggr]\;.
\eeeq

\begin{figure}[htb]
\vspace{9pt}
\begin{center}
\epsfig{figure=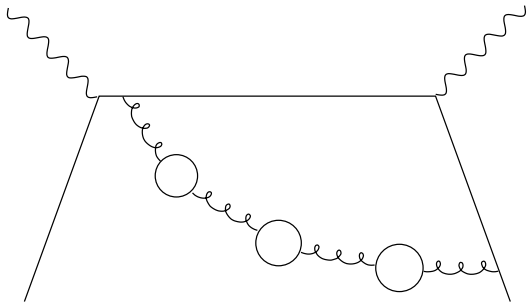,height=4.0cm}
\end{center}
\caption{Diagram containing a renormalon chain.}
\label{fig_chain}
\end{figure}

A known source of divergence is the class of diagrams
in which a chain of (quark) loops is inserted into
a gluon line as illustrated in \fref{fig_chain}.
They lead to a factorial divergence of the form
\beq
F = \sum_n c_n \as^n\;,\;\;\;\;\;c_n\sim n!\,(2b/p)^n
\eeq
where $b=-N_f/6\pi$ for $N_f$ flavours of quarks in
the loops and the number $p$, normally an integer,
depends on the observable $F$ being computed.

A crucial step in estimating the form of divergence
of the full perturbation series at high orders
is that of so-called `naive non-Abelianization'
\cite{BBNNA,Neu}, which
consists of making the replacement $N_f\to N_f-33/2$
 in the above equation, so that $b$ becomes
the first coefficient of the QCD $\beta$-function,
$b=(33-2N_f)/12\pi$. This takes account of gluon
loops and other contributions that give rise to the
running coupling. Surprisingly, the resulting estimates
of the perturbative coefficients $c_n$ seem to lie close
to the true values even when $N_f=0$ (pure gauge theory),
for quantities that have been studied to high orders
by means of lattice perturbation theory \cite{lattice}.

Assuming the approximate validity of the naive
non-Abelianization procedure, the divergent series
of contributions can be conveniently summarised
by the formula \cite{DokMarWeb} 
\beq\label{Fint}
F(Q^2,x) =\int\frac{d\mu^2}{\mu^2}\,
\aeff(\mu^2)\,\dot\cF(\mu^2/Q^2,x)\;.
\eeq
Here $Q^2$ is the hard process scale, assumed large,
and $x$ represents dimensionless ratios of other hard
scales, e.g. the Bjorken variable. As discussed more
fully in \cite{DokMarWeb}, $\mu^2$ is the variable of
integration in a dispersion relation but can be regarded
in some respects as an effective gluon mass-squared. The
`effective strong coupling' $\aeff$ is essentially
equivalent to $\as$ at large values of $\mu^2$ but
different from the perturbative expansion of $\as$ at
low scales. This discrepancy might or might not depend
on the observable $F$ -- we shall discuss later the
consequences of assuming that $\aeff$ is universal.
The `characteristic function' $\cF$ is found by
evaluating the observable $F$ to first order for
a finite gluon mass $\mu$. On dimensional grounds,
for $F$ dimensionless, this must be dependent on
$\mu$ only via the quantity $\eps=\mu^2/Q^2$.
$\dot\cF$ is then minus the logarithmic derivative
of $\cF$ with respect to $\eps$, i.e., with respect
to $\mu^2$.

\begin{figure}[htb]
\begin{center}
\epsfig{figure=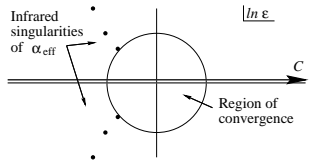,height=4.0cm,angle=270}
\end{center}
\caption{Integration contour and convergence region in the
$\ln\eps$ plane ($\eps=\mu^2/Q^2)$.}
\label{fig_plane}
\end{figure}

In general, $\dot\cF$ is a well-behaved function of
$\eps$, vanishing like a power at small $\eps$ and
an inverse power at large $\eps$ (modulo
logarithms). Thus we expect the integrand in
\eref{Fint}, considered as a function of the
complex variable $z=\ln\eps$, to be damped
exponentially in both directions along the
integration contour, which is the real axis
(\fref{fig_plane}).

To represent \eref{Fint} perturbatively, we
expand the effective coupling in terms of $\as(Q^2)$:
\beq
\aeff(\mu^2) = \aeff(e^z Q^2) = \sum_1^n a_n(z)
[\as(Q^2)]^n\;.
\eeq
This series has some finite region of convergence
in the $z$-plane, limited by the positions of the
infrared singularities of $\aeff$. If, for example,
we use the one-loop expression $\aeff(\mu^2)
= 1/b \ln(\mu^2/\Lambda^2)$, then the region
is a circle which touches the Landau pole at $z =
- \ln(Q^2/\Lambda^2)$. In this case a singularity
lies on the contour of integration and the integral
is not defined.  In a more realistic model for $\aeff$,
however, the infrared singularities will lie
elsewhere in the complex $z$ plane, as depicted in
\fref{fig_plane}, and the integral will be
unambiguous order by order in $\as(Q^2)$.
Nevertheless the perturbation series in $\as(Q^2)$
will diverge, because the contour extends outside
the region of convergence, both to the right and to
the left.  Since $\dot\cF$ is exponentially damped in
$z$ on both sides, while $a_n(z)$ is power-behaved,
with the power increasing with $n$, the divergence
will be of factorial form.

The divergence due to the segment of the contour
to the right of the convergence region is called an
{\em ultraviolet renormalon}.  We can see directly
from the integral representation (\ref{Fint}) that
this divergence is just an artefact of expanding in
powers of $\as(Q^2)$. The ultraviolet contribution
to the integral (say, that from $z>0$,
i.e.\ $\mu^2>Q^2$) is perfectly well
defined in terms of the usual perturbative form
of the running coupling $\aeff(\mu^2)\simeq \as(\mu^2)$.
It can be evaluated numerically, or equivalently
the corresponding series can be summed using a
standard technique such as Borel summation.
The divergence does not originate from the ultraviolet
region, being simply a reflection of the limit on the
region of convergence set by singularities far
away in the infrared.

On the other hand the divergence of the series due
to the region on the left, called an {\em infrared
renormalon}, cannot be eliminated  without
information on the non-perturbative behaviour
of $\aeff$ in the infrared region.  It should be clear
from \fref{fig_plane} that this ambiguity has
nothing to do with whether a Landau pole is
present on the axis \cite{DokUra}.
It will exist as long as $\aeff$
has any kind of infrared singularities, which it
surely must.  Therefore the so-called  infrared
renormalon ambiguity can only be resolved by
providing a physical model of the infrared
behaviour of $\aeff$.  Once this is done, one can
integrate unambiguously along the left-hand
portion of the contour, since no physical singularities
can lie on the $z$-axis (they would correspond to
tachyons).

\section{Power corrections}\label{sec_pow}
Assuming the validity of \eref{Fint}, we can
make useful deductions about the behaviour
of $F(Q^2,x)$ at low $Q^2$, without any specific
assumptions about the infrared behaviour of the
effective coupling. We shall argue that $F$
is of the form
\beq\label{Fsplit}
F(Q^2,x) = \Fptn(Q^2,x)+\Fpwn(Q^2,x)
\eeq
where the first term, representing the standard
$n$-th order perturbative prediction, has a
basically logarithmic $Q^2$-dependence while
the second, which includes non-perturbative effects,
is power-behaved.  Note that the second
term also requires a superscript $n$ since it depends
on where we truncate the perturbation series $\Fptn$.

To derive \eref{Fsplit} we split $\aeff$
into two parts, roughly described as perturbative and
non-perturbative:
\beq
\aeff(\mu^2) = \aptn(\mu^2) + \daen(\mu^2)\;.
\eeq
The superscript on $\aptn$ indicates that this
quantity is obtained by expanding to $n$-th order
in $\as(Q^2)$:
\beeq\label{aptn}
\aptn(\mu^2) &=&  \as(Q^2)+ b\left(\ln\frac{Q^2}{\mu^2}
+ K\right)  \as^2(Q^2) \nonumber\\
&& + \cdots+{\cal O}\left(\ln\frac{Q^2}{\mu^2}\right)^n
\as^n(Q^2)
\eeeq
where the constant $K$ depends on the choice of
renormalization scheme. Once again it follows that
the remainder $\daen$ also depends on $n$.

\begin{figure}[htb]
\begin{center}
\epsfig{figure=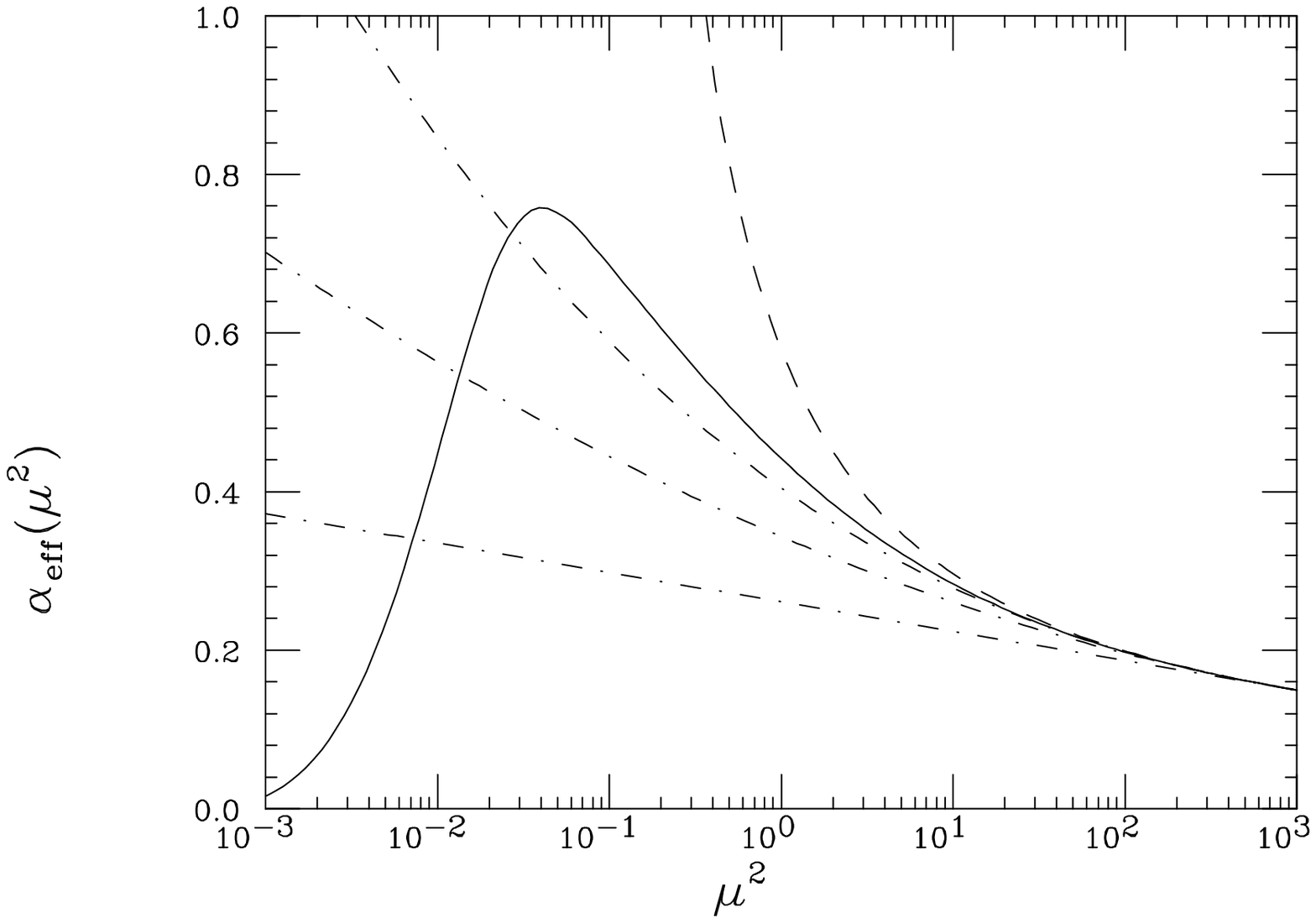,height=4.5cm}
\end{center}
\caption{Possible behaviour of the effective coupling.
Solid: $\aeff$. Dashed: 1-loop running $\as$. Dot-dashed:
$\aptn$ for $n=2,3,4$.}
\label{fig_aeff}
\end{figure}

In \fref{fig_aeff}, the solid curve shows a possible
model for the behaviour of $\aeff$, in which it is
assumed to vanish at very small scales, as suggested in
\cite{DoKhTr95}. Given the present level of understanding,
it could equally well approach a non-zero value. The
quantity $\daen$ is given by the difference between
the solid curve and the dot-dashed one representing
$\aptn$, which rises as $n$ increases. Therefore we
expect in general that $\daen$ will decrease with
increasing $n$.

Substituting in \eref{Fint}, we obtain \eref{Fsplit} with
\beq\label{Fpwn}
\Fpwn =\int_0^{\mI^2}\frac{d\mu^2}{\mu^2}\,
\daen(\mu^2)\,\dot\cF(\mu^2/Q^2,x)\;.
\eeq
We have assumed here that the approximation
(\ref{aptn}) to $\aeff$ is sufficiently accurate
at high $\mu^2$ for $\daen$ to be negligible
above some `infrared matching'
scale $\mI$ ($\Lambda\ll\mI\ll Q$).
It follows then that the $Q^2$ dependence of
$\Fpwn$ is determined by the behaviour of
$\dot\cF(\eps,x)$ for small values of
$\eps$. For the general behaviour
\beq\label{Flim}
\dot\cF(\eps,x) \sim f(x) \eps^{p/2} \ln^q \eps
\;\;\;\;\;\mbox{as}\;\;\eps\to 0
\eeq
we have
\beq\label{Fpwmom}
\Fpwn \sim \frac{f(x)}{Q^p}\int_0^{\mI^2}
\frac{d\mu^2}{\mu^2}\,\daen(\mu^2)\,\mu^p\,\ln^q
\left(\frac{\mu^2}{Q^2}\right)
\eeq
Thus, as stated earlier, $\Fpwn$ is power-behaved
(modulo logs), with $x$-dependence given by that of
$\dot\cF(\eps,x) $ at small $\eps$ and magnitude
determined by a (log-)moment of the `effective
coupling modification' $\daen$.

All this depends only on the validity of the
dispersive formula (\ref{Fint}), irrespective
of whether $\daen$ varies from one observable $F$
to another. Therefore one would expect the value of
the power $p$ to be given reliably by \eref{Flim}.
Under the additional assumption that $\aeff$,
and hence $\daen$, is universal, one obtains
testable relationships between power corrections
to different observables with the same value of $p$. 

Over the last few years, this `renormalon-inspired'
or `dispersive' approach to the estimation of power
corrections has been applied to a wide variety of QCD
observables \cite{DokMarWeb},\cite{Web94}-\cite{MeySchH1}.
Many of the results have been presented
in other talks at this meeting.  Therefore I shall
concentrate in the remainder of this talk on topics
that illustrate some of the general remarks made above.

\section{Structure functions}\label{sec_dis}

\begin{figure}[htb]
\begin{center}
\epsfig{figure=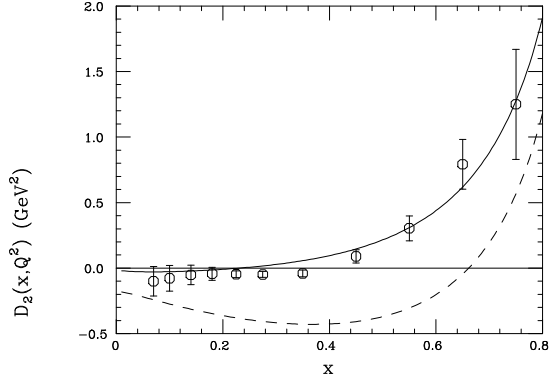,height=4.5cm}
\end{center}
\caption{Higher-twist contribution to $F_2$ (data, solid curve)
and $F_3$ (dashed curve).}
\label{fig_F2}
\end{figure}
\begin{figure}[htb]
\begin{center}
\epsfig{figure=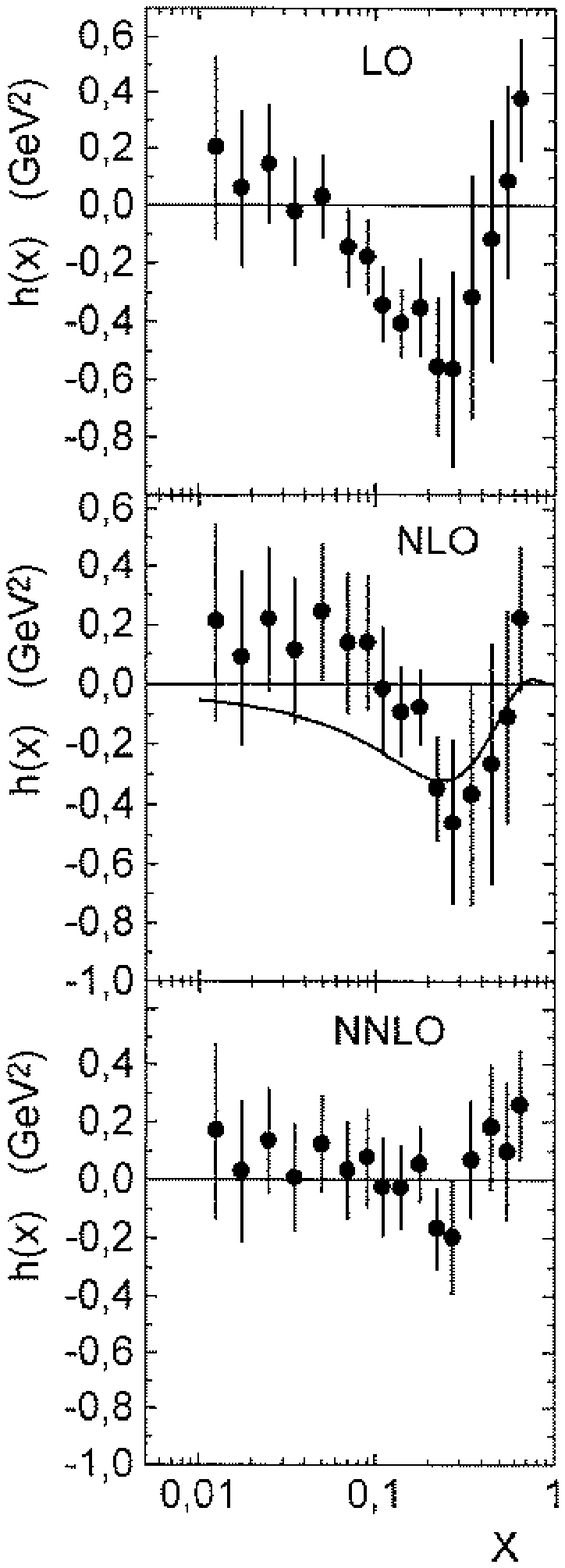,height=10.0cm}
\end{center}
\caption{Higher-twist contribution to $F_3$, extracted from
various orders of perturbative analysis. Figure adapted from
Ref.~[21].}
\label{fig_F3}
\end{figure}

The need for $1/Q^2$ power corrections to hadronic
structure functions, normally called higher-twist
contributions, is well established and their dependence
on the Bjorken $x$ variable has been extracted from
deep inelastic scattering (DIS) data.  Typical results are
shown in \frefs{fig_F2}{fig_F3} \cite{VirMil,KKPS}.
In \fref{fig_F2} the higher-twist contribution is
parametrized as $D_2(x)/Q^2$ time the leading-twist,
whereas in \fref{fig_F3} it is represented simply
as an additive term $h(x)/Q^2$.

Such contributions are predicted by the renormalon
approach, since the characteristic functions for the
evolution of (non-singlet) structure functions have the
general form
\beq\label{FDIS}
\dot\cF(\eps,x) \sim P(x) + f(x) \eps \ln\eps +\ldots
\eeq
where $P(x)$ is the usual splitting function and the
ellipsis represents terms that are less important as
$\eps\to 0$.  The convolution of the
(structure-function-dependent) coefficient function
$f(x)$ with the quark distribution then gives the
predicted form of the leading $1/Q^2$ correction.

The solid curve in \fref{fig_F2} shows the
prediction for the structure function $F_2$,
with the relevant ($p=2$, $q=1$) moment of $\daen$
(for $n=2$) in \eref{Fpwmom} as the only free
parameter \cite{DokMarWeb,DasWebDIS,Stein}.
The good agreement with the data is due mainly to
the characteristic $1/(1-x)$ behaviour at large $x$,
which is also predicted by a number of models
of higher-twist contributions.

A more stringent test is the prediction of the
power correction to $xF_3$ \cite{DasWebDIS},
which depends on the same moment
of $\daen$. The dashed curve shows the result,
for the value of the moment obtained by fitting $F_2$.
It is interesting that the prediction is negative for
all but the largest $x$ values, in agreement with
the need for a negative power correction to the GLS
sum rule (\ref{GLS}).  This is confirmed by a recent
analysis \cite{KKPS} of the data of the
CCFR collaboration \cite{CCFR}:  the
power corrections extracted in LO, NLO and
NNLO (i.e.\ $n=1,2,3$) analyses are shown in
\fref{fig_F3}, together with the same prediction
for $F_3$ as in \fref{fig_F2}.

\fref{fig_F3} also illustrates an important point
discussed in \sref{sec_pow}, namely that the normalization
of the power correction depends on the order of the
perturbative analysis. It decreases in magnitude
with increasing order, as more of the divergent
renormalon series is included in the perturbative
prediction $\Fptn$. We can see from the GLS
prediction, \eref{GLS}, that the perturbative
coefficients are indeed increasingly negative, consistent
with the presence of an infrared renormalon with a negative
coefficient, which tends to cancel the power correction.

\section{Event shape variables}\label{sec_shape}
Power corrections to hadronic event shapes are generally
larger than those in most other observables. This,
together with the lack of complete next-to-next-to-leading
order perturbative calculations of these quantities, has made
it difficult to use them for precise tests of QCD, in spite
of the very high statistics available from LEP1.

Usually the non-perturbative contributions to event shapes
are estimated using Monte Carlo event generators \cite{JS,HW},
or rather the hadronization models that are built into them.
The difference between the parton- and hadron-level predictions
of the generator is taken as the hadronization correction,
with an error given by the discrepancy between the models.
As discussed in \sref{sec_intro}, the trouble
with this is that the parton-level generators
do not coincide with fixed-order perturbation theory.
They represent an approximate summation of various
contributions (including the renormalon graphs) up to
rather high order. In addition, they contain cutoff
parameters that can emulate non-perturbative effects.
Thus the power corrections might be under- or over-estimated
by this procedure, even though the generator might agree
perfectly with the data at the hadron level.
 
Some less model-dependent insight into power corrections to event
shapes can be gained from the renormalon-inspired type of treatment
discussed above. Consider as a prototype the thrust variable
in $\ee$ annihilation,
\beq
T = \max \sum_i |\bm{p}_i\cdot\bm{n}|/\sum_i |\bm{p}_i|\;,
\eeq
where the sum is over final-state hadron momenta $\bm{p}_i$
and the maximum is with respect to the direction of the unit
vector $\bm n$ (the thrust axis). In lowest order the difference
$1-T$ arises from a $q\bar q g$ final state. Making a
Sudakov decomposition of the gluon momentum $k$,
\beq
k = \alpha p + \beta \bar p +k_\perp
\eeq
where $p^2 =\bar p^2=p\cdot k_\perp=\bar p\cdot k_\perp=0$,
$k_\perp^2 = -\kt2$ and $2p\cdot\bar p = Q^2$,
the c.m.\ energy-squared, we have
\beq\label{talpbet}
t\equiv 1-T = \min\{\alpha,\beta\}
\eeq
in the soft region. Hence the characteristic function
for the mean value of this quantity is
\beeq
\cF &\simeq& \frac{C_F}{\pi}\int\frac{d\alpha}{\alpha}
\frac{d\beta}{\beta}\,d\kt2\nonumber \\
& &\cdot\min\{\alpha,\beta\}\,\delta(\alpha\beta Q^2
-\kt2-\mu^2)\nonumber \\
&=&\frac{2C_F}{\pi Q}\int\frac{d\kt2}{\sqrt{\kt2+\mu^2}}\;,
\eeeq
from which we find
\beq\label{dotFt}
\dot\cF \simeq \frac{2C_F}{\pi} \frac{\mu}{Q}\;.
\eeq
It follows from \erefs{Fpwn}{dotFt} that the quantity
$\VEV{t}$ is expected to receive a $1/Q$ power
correction, in contrast to the $1/Q^2$ corrections
to other quantities like structure and fragmentation
\cite{frag} functions. A similar result is obtained for
the mean values of several other $\ee$ shape
variables \cite{Web94,AkZak,KorSte},
and for the corresponding quantities
in deep inelastic final states \cite{DasWebshap}.

There is by now good experimental evidence for $1/Q$ corrections
to the mean values of several event shape variables in both $\ee$
annihilation \cite{DELPHI,JADEOPAL} and DIS \cite{H1}. Furthermore
the magnitudes of the corrections are in most cases consistent
(at about the 20\% level) with a universal value of the relevant
moment of $\dae2$ in \eref{Fpwmom}, namely
\beq\label{daemom}
\int \frac{d\mu^2}{\mu^2}\,\mu\,\dae2(\mu^2)\simeq 1\;\mbox{GeV}\;.
\eeq

Turning from the mean values to the distributions of
event shape variables, a simple result is obtained
when the dominant low-momentum contributions
{\em exponentiate}, that is, when they correspond to
essentially uncorrelated multiple soft gluon emission.
This is the case for the thrust distribution, which
may be expressed as a Laplace transform \cite{CTTW}
\beq
F(t)\equiv \frac{1}{\sigma}\frac{d\sigma}{dt}
= \int_C d\nu\,\exp[\nu t + R(\nu,Q)]
\eeq
where the contour $C$ runs parallel to the imaginary axis,
to the right of all singularities of the integrand, and $t=1-T$
as usual. The `radiator' $R$ has the following form in lowest order:
\beeq
R(\nu,Q) &=& \frac{2C_F}{\pi} \int\frac{d\mu^2}{\mu^2}\aeff(\mu^2)
\nonumber \\
& & \cdot\int_{\mu^2/Q^2}^{\mu/Q}\frac{du}{u}
\left(\mbox{e}^{-u\nu}-1\right)
\eeeq
To compute the effect of a non-perturbative contribution to
$\aeff$, confined to the region $\mu^2<\mI^2\ll Q^2$, we note that,
since $\nu$ is conjugate to $t$, the exponential can safely be
expanded to first order as long as $t\gg \mI/Q$. The resulting
change in the radiator is
\beq
\delta R \simeq -\frac{2C_F}{\pi}\nu \int\frac{d\mu^2}{\mu^2}\daen(\mu^2)
\frac{\mu}{Q}\equiv -\nu\frac{A}{Q}
\eeq
which corresponds simply to a {\em shift} in the
$t$-distribution \cite{KorSte,DokWeb97}, by an
amount $A/Q$ equal to the non-perturbative shift
in $\VEV{t}$ computed above:
\beq\label{Ftshift}
F(t)\simeq\Fper(t-A/Q)
\eeq
where $\Fper$ is the perturbative prediction.
This agrees well with the $\ee$ data \cite{DokWeb97}.

There is some ambiguity in the calculations outlined above:
we assumed an effective gluon mass $\mu$ but used the massless
forms of the shape variable and lowest-order matrix element,
when expressed in terms of the Sudakov variables
$\alpha$ and $\beta$, as was done in \cite{Web94}.
In reality the `massive' gluon fragments into
massless partons, which eventually form hadrons,
and the fragmentation products may have a thrust value
different from that of their parent \cite{NaSey}. This
will not change the $1/Q$ form of the power correction
but it would be expected to modify its coefficient.

These issues have been clarified in a recent paper
from the Milan group \cite{Milan}, in which the coefficient
of the $1/Q$ correction is evaluated to two-loop order
in the soft limit. A substantial increase, relative to
the above naive treatment, is
obtained -- a factor of 1.8 for three active flavours.
Interestingly, however, this enhancement factor itself
appears to be universal across a wide range of shape
variables, in both $\ee$ and DIS final
states \cite{inprep}.  Therefore the experimental
analyses \cite{DELPHI,JADEOPAL,H1} of event shapes
assuming universality are not invalidated,
but the fitted values of the moments of $\daen$,
such as \eref{daemom}, need to be rescaled by the
inverse of the Milan factor.

\section{Better shape variables}\label{sec_better}
As we have seen, the power corrections to event shapes
are normally of order $1/Q$.  For determinations of
$\as$, it would be an advantage to know if there are
any shape variables that are free of such corrections,
preferably up to order $1/Q^2$.

One set of variables that has been suggested is the
higher moments of shape variables. A straightforward
approach along the lines of \sref{sec_pow}
suggests that if the mean value has a correction of
order $1/Q$ then the $p$-th moment will only receive
a non-perturbative contribution of order $1/Q^p$.

Unfortunately the situation is not so simple, since we
would also like to safeguard against corrections of order
$\as(Q^2)/Q\gg 1/Q^2$.  Consider again for example the variable
$t\equiv 1-T$.  We know that the power correction to
the NLO perturbative prediction for $\VEV{t}$ is $A/Q$
where $A\simeq 1$ GeV. Furthermore we saw that the
main non-perturbative effect on the thrust distribution
$F(t)$ is a simple shift, \eref{Ftshift}. It follows that
for the mean value of $t^p$ we have
\beeq\label{VEVtp}
\VEV{t^p} &=& \int dt\,t^p F(t)\simeq
%\int dt\,t^p\Fper(t-A/Q)\nonumber\\ &=&
\int dt\,(t+A/Q)^p\Fper(t)\nonumber\\
&=&\stub{\VEV{t^p}}{pert} + \frac{pA}{Q}
\stub{\VEV{t^{p-1}}}{pert}
+ {\cal O}\left(\frac{1}{Q^2}\right)
\eeeq
In practice, the perturbative predictions for
$\VEV{t^p}$ decrease rapidly with $p$, and therefore
the second term is relatively important for $p>1$,
in spite of its suppression by $\as(Q^2)$. One finds
\beeq
\stub{\VEV{t}}{pert} &=& 0.335\as + 1.03\as^2 +\ldots\;, \nonumber \\
\stub{\VEV{t^2}}{pert} &=& 0.030\as + 0.15\as^2 +\ldots\;.
\eeeq
and so, at LEP energies and below, the power corrections to $\VEV{t}$
and $\VEV{t^2}$ are actually comparable, relative to the perturbative
predictions. In addition, the perturbation series for $\VEV{t^2}$
and higher moments show increasingly poor convergence.

A better idea is to consider moments about the mean value,
such as the {\em thrust variance} $\stub{\sigma^2}{T}
= \VEV{t^2}-\VEV{t}^2$. Then we see from \eref{VEVtp} that
the $\as(Q^2)/Q$ corrections should cancel out. Furthermore
the perturbative prediction,
\beq
\stub{\sigma^2}{T, pert} = 0.030\as + 0.04\as^2 +\ldots \;,
\eeq
shows better convergence than that for $\VEV{t^2}$ -- better
than $\VEV{t}$, in fact. Nevertheless,
this is a quantity an order of magnitude smaller than $\VEV{t}$,
and so its measurement will require that experimental sources of
error are well under control.

An interesting open question is whether the $1/Q$ and $\as(Q^2)/Q$
corrections to $\stub{\sigma^2}{T}$ remain cancelled after taking
into account the Milan enhancement factor, to be extracted from a
two-loop analysis.

Finally we should consider a shape variable which is not expected
to have $1/Q$ or $\as(Q^2)/Q$ corrections to its first moment,
namely the three-jet resolution $y_3$. This is the value of the
Durham jet measure $\stub{y}{cut}$ at which three jets are just
resolved in the final state.  Recalling that the Durham
criterion for resolving two partons $i$ and $j$ is
$y_{ij}>\stub{y}{cut}$, where \cite{Durham} 
\beq
y_{ij} = 2\min\{E_i^2,E_j^2\}(1-\cos\theta_{ij})/Q^2\;,
\eeq
we find that in the soft limit we obtain in place of \eref{talpbet}
\beq\label{yalpbet}
y_3 = (\alpha+\beta)\min\{\alpha,\beta\}\;,
\eeq
giving instead of \eref{dotFt}
\beq\label{dotFy}
\dot\cF \simeq \frac{2C_F}{\pi} \frac{\mu^2}{Q^2}
\ln\left(\frac{Q}{\mu}\right)
\eeq
for $\VEV{y_3}$. Thus, modulo possible logarithms, the expected
leading correction to this quantity is ${\cal O}(1/Q^2)$.
Once again, it would be interesting to know whether the Milan
enhancement can somehow generate a $1/Q$ correction.
Experimentally, no significant power correction to $\VEV{y_3}$
has yet been found \cite{JADEOPAL},
which is certainly consistent with the absence of a $1/Q$ term.

\section{Conclusions}\label{sec_conc}
The renormalon-inspired or dispersive method for estimating
power corrections to perturbative calculations of QCD
observables has proved to be a useful phenomenological
tool. At the very least, it warns us of those quantities
and phase-space regions in which the perturbative
calculation is dangerously sensitive to low-momentum
contributions. If we wish, we can simply avoid those
regions in comparing theory with experiment. From this
viewpoint, the renormalon ambiguity is analogous to
the renormalization-scale dependence of a next-to-leading
order perturbative prediction: if it is large, we would be
unwise to trust the prediction as a reliable test of QCD
or as a good way of measuring $\as$. The ideal observable
for these purposes has both a small scale dependence
and a small renormalon ambiguity -- in particular, a
large value of the power $p$. These are somewhat independent
requirements, and therefore a renormalon analysis provides
useful additional information, even if we do not accept it
as a quantitative estimate of power corrections. 

As far as I am aware, the powers $p$ for the corrections
observed experimentally agree with those predicted in all
cases studied so far. This provides the motivation for going
further and comparing the measured and computed coefficients
of power corrections, assuming a universal low-energy form
for the effective strong coupling $\aeff$. The initial
results look encouraging to me, although many issues
remain unclear.  As emphasised in \sref{sec_dis}, the
coefficient must depend on the order of the perturbative
prediction that is being subtracted. Presumably the
optimal comparison would be with the perturbation
series truncated at its smallest term, but we remain
far from that situation. In the case of
event shape variables, the renormalization of the
coefficient by two-loop effects has been shown to be
large but appears to remain universal for a range
of observables. A general argument can be made that
higher loop corrections should be small \cite{Milan},
but explicit calculations would be reassuring.

Another relevant question
is whether we should expect the strong coupling itself
to exhibit power corrections, possibly as large as
$1/Q^2$ \cite{AkZak97}. There have been recent suggestions
of such terms in lattice calculations \cite{lat97}.
A systematic perturbative treatment of the effective strong
coupling beyond one-loop (i.e.\ one-renormalon-chain) order
seems essential for this purpose, but is still lacking.
Approaches such as that of \cite{Wat97} may be fruitful.
Such a treatment would also be valuable for the study of
power corrections to gluon-dominated quantities such as
hadronic jet shapes \cite{Sey97} and small-$x$ structure
functions \cite{Haut}.

\section*{Acknowledgments}
It is a pleasure to thank the organizers for arranging such a
stimulating meeting. I am grateful to many colleagues, especially
M.\ Dasgupta, Yu.L.\ Dokshitzer, G.\ Marchesini, P.\ Nason,
G.P.\ Salam and M.H.\ Seymour, for helpful discussions.

\end{document}